\documentclass[%
 reprint,
 preprintnumbers,
 amsmath,amssymb,
 aps,onecolumn,
 nofootinbib
]{revtex4-1}

\usepackage{graphicx}
\usepackage{dcolumn}
\usepackage{bm}

\newcommand{\sL}{\!\scriptscriptstyle L}
\newcommand{\sR}{\!\scriptscriptstyle R}

\newcommand{\PP}{\mathbb{P}}
\newcommand{\PL}{\PP_{\sL}}
\newcommand{\PR}{\PP_{\sR}}

\newcommand{\Slash}{\slash\!\!\!}

\newcommand{\sbar}{\overline{s}}

\begin{document}

\title{\boldmath Dark $Z$ Implication for Flavor Physics}

\author{Fanrong Xu$^{a,b,c}$}
\affiliation{
$^{a}$Institute of Physics, Academia Sinica,\\
Taipei, Taiwan 11529, R. O. China\\
$^{b}$Department of Physics, Jinan University,\\
Guangzhou 510632, P.R. China\\
$^{c}$Kavli Institute for Theoretical Physics China, Chinese Academy of Sciences\\
 Beijing 100190, P. R. China}


\begin{abstract}
 Dark $Z$/dark photon ($Z'$) is one candidate of dark force carrier, which helps to
 interpret the properties of dark matter (DM). 
  Other than conventional  studies of DM including direct detection, indirect detection and 
 collider simulation, in this work we take flavor physics as a complementary approach to
 investigate the features of dark matter. 
We  give an exact calculation of the new type of penguin diagram induced by $Z'$ which
 further modifies the well-known  $X, Y, Z$ functions in penguin-box expansion.
 The measurement of rare decays $B\to K^{(*)}\mu^+\mu^-$ and $B_s\to \mu^+\mu^-$ at LHC,
 together with direct CP violation $\varepsilon'/\varepsilon$ in $K\to\pi \pi$ as well as $K_L\to\mu^+\mu^-$,
 are used to determine the parameter space. The size of coupling constant, however,
 is  found to be $\mathcal{O}(1)$ which is much weaker than the known
 constraints.
 \end{abstract}


\maketitle

\section{Introduction}
\label{sec:intro}

Dark matter (DM) constitutes about $27\%$ of the  energy-matter budget of universe,
significantly more than $5\%$ of baryonic matter\cite{Ade:2013ktc}.
Nevertheless, the exact nature of DM is kept mysterious so far.
Yet it is still unclear whether DM can be described by scalar, fermion, vector or even graviton.

Among various DM candidates, dark photon  has been the one of particular interest.
The idea was initialed in 1980s \cite{Holdom:1985ag}, 
and developed in recent years \cite{dev}\cite{Davoudiasl:2012ag}.
Suppose there is an extra $U(1)_D$ group besides SM gauge group,
under which  
all the SM interactions are invariant. 
The gauge boson of $U(1)_D$, named dark photon,
interacts with SM  $U(1)_Y$ gauge boson
via a kinetic mixing term.  
It helps to explain the astrophysical observation of positron excesses \cite{Adriani:2008zr},
as well as other astrophysical phenomenology such as supernova bounds \cite{Kazanas:2014mca} and Big Bang
Nucleosynthesis\cite{Foot:2014uba}.  
The direct search of dark photon, for example XENON 100, has put a very strong constraint \cite{An:2014twa}.
For $e^+e^-$ collider, a recent search performed at BaBar shows
a null result, neither finds nor rules out dark photon \cite{Lees:2014xha}.
A further experiment, the Heavy Photon Search (HPS) experiment located at Jefferson Lab \cite{Moreno:2013mja},
is designed  to search dark photon in the mass range $20\,\mathrm{MeV}$ to $1\,\mathrm{GeV}$
as well as the related coupling.
However, it has been realized recently that a simple dark photon is not favored 
by $3.6 \sigma$ deviation of muon anomalous magnetic moment \cite{Adare:2014mgk}.
Given the fact that dark photon model is the extreme case when the parameter to describe 
$Z$-$Z'$ mass mixing is closed in a more generic dark $Z$ model, 
it is necessary to extend dark photon to dark $Z$  which is the working frame 
of this paper. 

Flavor physics is not only taken as a platform for precise test of SM, but also plays an important role
in indirect search of new physics (NP) beyond Standard Model (SM). 
Great progresses have already been made since LHC runs. For example,
it was hoped for decades that NP might exist in zero crossing point 
$q_0^2$ of the differential branching ratio of
$B\to K^*\mu^+\mu^-$ (see for example \cite{Hou:2011fw})  
but finally turned out tiny NP effect\cite{Aaij:2013iag}. 
Taken DM theory as one kind of ordinary NP theory, flavor physics then 
would provide as a complementary way
for conventional approaches of DM study, including direct detection, indirect detection
and collider production. 
Based on the great achievement made in Run I, 
the LHC has already started its Run II in 2015.
Then it would be interesting, and timely,
to connect together these
two different fields, dark matter and flavor physics. A similar effort can also be found in 
\cite{Sierra:2015fma}.

The paper is organized as follows. In section 2, we will briefly set up the dark $Z$ model.
An exact result of $Z'$ penguin and further modifications to $X, Y, Z$ functions are given
in Section 3.  Some typical processes which are affected by the $Z'$ 
including the recently measured $B\to K^*\mu^+\mu^-$ and
 $B_s\to \mu^+\mu^-$, as well as $\varepsilon'/\varepsilon$ and $K_L\to \mu^+ \mu^-$ 
 are discussed in section 4, also the relevant formulas are given therein.
 In section 5 the obtained numerical results are shown, based on which 
 we will make a discussion. The details of $Z'$ penguin calculation can be found in appendix.

\section{The model}
\label{sec:model}

Suppose there exits an extra $U(1)$ group, other than SM $U(1)$,  what it brings in phenomenology
 is an interesting question.
It was considered how electromagnetic charge is shifted by this extra $U(1)$ group in the initial paper
 \cite{Holdom:1985ag}. Until recently it becomes popular to take this $U(1)$ gauge boson as 
 a DM candidate.
 
Under the  dark group, notated as $U(1)_D$,
all the SM interactions are invariant .
The connection between dark photon with SM particles is from a kinetic mixing term, leading to
the effective Lagrangian \cite{Davoudiasl:2012ag}
\begin{equation}
\mathcal{L}=-\frac14 \hat{B}_{\mu\nu}\hat{B}^{\mu\nu}+\frac12 \frac{\epsilon}{\cos\theta_W}
\hat{B}_{\mu\nu}\hat{Z}^{'\mu\nu}
-\frac14\hat{Z}'_{\mu\nu}\hat{Z}^{'\mu\nu}, 
\end{equation}
where $\theta_W$ is Weinberg angle,  $\hat{Z}'$  and $\hat{B}$ are dark photon and SM $B$ field with
the corresponding field strength
\begin{equation}
\hat{B}_{\mu\nu}=\partial_\mu \hat{B}_\nu -\partial_\nu \hat{B}_\mu,\qquad
\hat{Z}'_{\mu\nu}=\partial_\mu \hat{Z}'_\nu -\partial_\nu \hat{Z}'_\mu.
\end{equation}
and the mixing of gauge bosons is mimicked by parameter $\epsilon$, which
is supposed to be small and need to be determined.
The convention above, in gauge interaction state, is not diagonalized. 
By redefining  fields as,
\begin{equation}
\left(\begin{array}{c} Z'_0\\ B\end{array}\right)=\left(\begin{array}{cc}
\sqrt{1-\frac{\epsilon^2}{c_W^2}}&0\\
-\frac{\epsilon}{c_W}&1\end{array}\right)
\left(\begin{array}{c} \hat{Z'}\\ \hat{B}\end{array}\right)\label{eq:rot1}
\end{equation}
the Lagrangian is then rotated to a diagonal form
\begin{equation}
\mathcal{L}=-\frac{1}{4}B_{\mu\nu}B^{\mu\nu}-\frac{1}{4}Z'_{0,\mu\nu}Z_0^{'\mu\nu}.
\end{equation}
Note here the field after rotation with a subscript $0$ differs from the one before rotation with a hat.
In SM the $B$ field can be projected to
photon and $Z$ after spontaneous symmetry breaking (SSB). 
Incorporating $Z'_0$, the
related neutral gauge fields are shifted,
\begin{align}
&A=\hat{A}-\epsilon\hat{Z}'_0\nonumber\\
&Z_0=\hat{Z}_0+\epsilon \tan\theta_W\hat{Z}_0\nonumber\\
&Z'_0=\hat{Z}'_0.
\end{align}
The rotation does not change the definition of $Z'_0$,
however, photon and $Z$ field  are modified indeed. 
Due to this modification of gauge fields, the interaction
between $Z'_0$ and ordinary matter is induced, which is named
as dark photon model.

When $Z'_0$-$Z_0$ mass mixing is considered, the simple
dark photon model is then extended to dark $Z$ model. 
Generally speaking, the mass of $Z'$ could either be added by hand
which is called St\"{u}ckelberg mechanism\cite{Ruegg:2003ps} (the origin of 
St\"{u}ckelberg photon, for example, is discussed in string theory\cite{Feng:2014cla}) 
or by applying Higgs mechanism, see  \cite{Davoudiasl:2012ag} as an example.
In this paper, we shall adopt the treatment of $Z'$ mass in the latter case,
without involving the details of the mechanism itself. 
After the neutral gauge bosons obtain mass after SSB,
a further
rotation is required after the one in eq.(\ref{eq:rot1}) for diagonalising 
mass matrix
\begin{equation}
\left(\begin{array}{c} Z\\ Z'\end{array}\right)
=\left(\begin{array}{cc}\cos \zeta & -\sin \zeta\\
\sin\zeta & \cos\zeta\end{array}\right)\left(\begin{array}{c}
Z_0\\Z'_0\end{array}\right)\label{eq:rot2}
\end{equation}
where the rotation angle $\zeta$ is model dependent and  analytically 
might be complicated, but numerically should be small, 
(for example, see \cite{Davoudiasl:2012ag}). 
Now combine together the two rotations eq. (\ref{eq:rot1}) and eq. (\ref{eq:rot2}),
the modifications to photon and $Z$ by dark $Z$ shows 
\begin{subequations}
\begin{align}
& A_\mu=\hat{A}_\mu-\epsilon Z'_\mu\\
& Z_\mu=\cos\zeta \hat{Z}_{0,\mu}-\epsilon_Z{Z}'_{\mu}\approx \hat{Z}_{0,\mu}-\epsilon_Z{Z}'_{\mu}.
\end{align}
\end{subequations}
Formally the shift of neutral fields in dark photon model
is characterised by two independent parameters
$\epsilon$ and $\epsilon_Z$, respectively.
In fact $\epsilon_Z$ also has a lengthy analytical expression based on detailed model.
The equivalent $\epsilon_Z$ defined to replace the  complicated structure
brings the convenience. The dark $Z$ field here and also hereafter is denoted as $Z'$, with
original SM field denoted with a hat. 
The two coupling constants together with $Z'$ mass constitute the unique $3$
model parameters of dark Z, which 
could  be measured in experiments. 

Apparently the interactions between $Z'$ and SM particles are simply induced by the 
shifted neutral gauge field.  Explicitly, we show how $Z'$ couples to SM fermions, 
\begin{equation}
\mathcal{L}_{Z'ff}=-\left(\epsilon e J_{em}^\mu+ \epsilon_Z\frac{g}{2\cos\theta_W}J_{NC}^\mu\right)Z'_\mu
\end{equation}
in which the SM electrical current and weak neutral current are
\begin{align}
&J_{em}^\mu= Q_f\bar{f}\gamma^\mu f,\qquad\nonumber\\
&J_{NC}^\mu=(T_{3f}-2Q_f\sin^2\theta_W)\bar{f}\gamma^\mu f -T_{3f}\bar{f}\gamma^\mu\gamma_5 f,
\end{align}
where $f$ stands for fermions with corresponding electric charge $Q_f$, isospin $T_{3f}=\pm \frac12$.  
With both vector coupling and axial-vector coupling, $Z'$ 
behaves as a light version of $Z$ 
and heavier version of photon.
For the coupling of $Z'$ and other gauge bosons, it has both ``$Z$ component'' and ``$A$ component''
sized by $\epsilon$ and $\epsilon_Z$ respectively.

\section{The $Z'$ effect in FCNC processes}

As current energy frontier, the LHC brings plentiful opportunities for flavour physics which
dominated by flavour changing neutral current
(FCNC) processes.
In this section, we will investigate these processes of meson physics in dark $Z$ model.
It is known in SM  FCNC processes are induced at loop level.
Conventionally the Feynman diagrams contributing
 to FCNC can be classified to three point penguin diagram and four point box diagram. 
 If NP exists, the new interaction brought in by NP
 will modify parts/all of these SM penguin and box contribution. Within the $Z'$ model working frame,
however, this modification is only applied in photon penguin and $Z$ penguin, keeping
box diagram contribution unchanged. 
To make the new effect more distinguishable, we extract these modifications alone
and name it as  $Z'$-penguin specifically.

\subsection{$Z'$ penguin}
\label{sec:Zprime}

We take $b\to s Z'$ as an example, noting similar result can be applied to $b\to d Z'$ and $s\to d Z'$
when necessary conditions are satisfied.

\begin{figure}[h!]
\begin{center}
{
 \includegraphics[width=100mm]{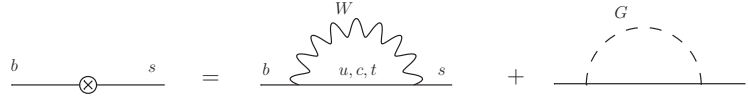}
}
\end{center}
\vskip0.55cm
\caption{
The effective vertex for $b\to s$ transition.}\label{bsr:wf}
\end{figure}

\begin{figure}[h!]
\begin{center}
{
 \includegraphics[width=150mm]{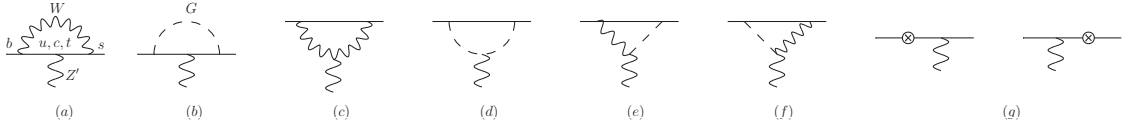}
}
\end{center}
\vskip0.55cm
\caption{
The Feynman diagrams contributing to $b\to s Z'$ in Feynman-t' Hooft gauge. Fig. $(a)$ to $(f)$ are three point diagrams while
Fig. $(g)$ is from the correction to external leg where the effective vertex denoted by a cross is explained
in Fig.\ref{bsr:wf}. }\label{bsr:vertex}
\end{figure}

In Feynman-t' Hooft gauge there are totally $10$ Feynman diagrams giving contributions to $b\to s Z'$.
During our realistic calculation, we group two of the external leg corrections and
replace them by an effective vertex shown in Fig. \ref{bsr:wf},
with $Z'$ inserted in either two legs. 
The four self-energy diagrams, effectively two, are then shown 
as $(g)$ in Fig. \ref{bsr:vertex}, together with the remaining six
ordinary three point diagrams given as $(a)$ to $(f)$.

As mentioned above the full result can be decomposed into ``$A$ component'' (or $\epsilon$ component) 
and ``$Z$ component''
(or $\epsilon_Z$ component). The $\epsilon$ component is same as $b\to s\gamma$ while the latter one
is similar to $b\to s Z$. For photon penguin, there are two types of effective verteices corresponding to 
real and virtual photon. In below since we focus on semileptonic processes thus only the virtual
photon vertex is taken into account. In order to keep the final result the same structure
as photon penguin, the $\epsilon_Z$ component differs from SM $Z$ penguin by neglecting 
dipole term contribution.  An exact calculation, shown in appendix \ref{sec:appendix},  gives the $Z'$ penguin vertex as
\begin{equation}
\sbar\Gamma^\mu b\big|_{Z'}=  i\lambda_i \frac{G_F}{\sqrt{2}}\frac{e}{8\pi^2}H_0(x_i)\sbar
 (q^2\gamma^\mu- q^\mu\Slash q)(1-\gamma_5) b,
\end{equation}
in which 
$\lambda_i=V_{ib}V^*_{is}$ $(i=u, c, t)$, $x_i=\frac{m_i^2}{m_W^2}$ and
$q$ is outgoing momentum carried by gauge boson. 
The vertex function $H_0(x_i)$ consisting of photon component function  $ {D}_0(x_i)$  and  newly calculated 
$Z$ component function $\tilde {D}_0(x_i)$,
are characterised by $\epsilon$ and $\epsilon_Z$, giving
\begin{subequations}
\begin{align}
&H_0(x)=\epsilon D_0(x)+\epsilon_Z \tilde{D}_0(x)\\
&D_0(x)=-\frac49 \ln x + \frac{-19 x^3+25 x^2}{36(x-1)^3}+\frac{x^2(5x^2-2x-6)}{18(x-1)^4}\ln x\\
&\tilde{D}_0(x)=
-\frac{1}{s_Wc_W}\left[ \frac{34x^3-141x^2+147x-58}{216(-1)^3}
+\frac{(-3x^4+18x^3-27x^2+19x-4)\ln x}{36(x-1)^4}\right.\nonumber\\
&\hspace{1.5cm} 
+ \left.c_W^2\left( \frac{-47x^3+237x^2-312x +104}{108(x-1)^3}
+\frac{(3x^4-30x^3+54 x^2-32 x+8)\ln x}{18(x-1)^4}\right)\right],
\end{align}
\end{subequations}
with $s_W=\sin\theta_W, c_W=\cos\theta_W$.
Note during evaluating, the light down type quark is supposed to be massless and 
the momentum transfer is small, comparable to light quark mass. 
We also assume the light $Z'$ mass smaller than the mass threshold for muon 
pair production, which guarantees no $Z'$ resonance is produced when 
final state of charged lepton is muon. Also in this work we will not touch
electron and neutrino as the lepton final state for the production of $Z'$
when $m_{Z'}$ is above electron threshold. Nevertheless, it is safe to neglect
$m_{Z'}$ in this paper.

\subsection{The modification to $X, Y, Z$ function}

In the state-of-the-art effective Hamiltonian approach, physical observables can 
be factorized into short distance (SD) and long distance (LD) contribution.
The SD part is treated in perturbative theory while LD hadronic matrix 
resorts to various methods including lattice QCD.
The SD contribution gives various combination of penguin diagrams and box diagrams and
leads to he so-called $X, Y, Z$ functions, which also
depends on both the theoretical working frame and the calculated physical observable.
For example, $Y$ is to characterize the Wilson coefficient in the 
effective Hamiltonian of $B_q\to \ell^+\ell^-$, and for a more complicated process
$B\to K^*\mu^+\mu^-$  more functions are involved.
In the frame of $Z'$ model, we have already discussed a new $Z'$ penguin in previous sector.
Before combining detailed phenomenology,
to include $Z'$ contribution in standard $X, Y, Z$ functions systematically is now our target, 

In SM the $X, Y, Z$ functions are obtained via combing 
$bs\gamma$, $bsZ$, box diagram vertex (see  \cite{Buras:1998raa}), and now 
we need to contain $bsZ'$ vertex.
The amplitudes of $b\to s \bar{\ell}\ell$ (or $\sbar  b \to \bar{\ell} \ell$) 
mediated by different gauge bosons are in the form of
\begin{align}
&i\mathcal{M}_{\gamma}=i\lambda_i \frac{G_F}{\sqrt{2}}\frac{\alpha}{2\pi} D_0(x_i) (\sbar b)_{V-A} (\bar{\ell} \ell)_V\\
&i\mathcal{M}_Z= i\lambda_i \frac{G_F}{\sqrt{2}}\frac{\alpha}{2\pi s_W^2} 2 C_0(x_i)
\left[v_f(\sbar b)_{V-A} (\bar{L} L)_V -a_f(\sbar b)_{V-A} (\bar{L} L)_A\right]\nonumber\\
&i\mathcal{M}_{Z'}= i\lambda_i \frac{G_F}{\sqrt{2}}\frac{\alpha}{2\pi s_W^2}2C_0(x_i)\left[
\frac{s_WH_0(x_i)}{4c_WC_0(x_i)}v'_f (\sbar b)_{V-A}(\bar{L}L)_V
-\frac{s_WH_0(x_i)}{4c_WC_0(x_i)}a'_f (\sbar b)_{V-A}(\bar{L}L)_A
\right]\nonumber
\end{align}
with $L=(\ell, \nu)$ and the coupling of $Zff$ and $Z'ff$
\begin{align}
&v_f=T_{3f}-2Q_f s_W^2,\qquad
a_f=T_{3f},\nonumber\\
&v'_f= \epsilon\cdot 2 Q_fs_W c_W + \epsilon_Z (T_{3f}-2Q_f s_W^2)
,\qquad a'_f=\epsilon_Z T_{3f}.
\end{align}
For convenience $Z$ and $Z'$ contribution can be put together in a 
compact form by showing
exact final state 
\begin{align}
& i\mathcal{M}_{ZZ'}^{\bar{\ell}\ell}= i\lambda_i\frac{G_F}{\sqrt{2}}
\frac{\alpha}{2\pi s_W^2} C_0(x_i)\left[(-1+4 s_W^2)(1+\delta_1)(\sbar b)_{V-A}(\bar{\ell}\ell)_V
+(1+\delta_2)(\sbar b)_{V-A} (\bar{\ell} \ell)_A\right]\nonumber\\
& i\mathcal{M}_{ZZ'}^{\bar{\nu}\nu}= i\lambda_i\frac{G_F}{\sqrt{2}}
\frac{\alpha}{2\pi s_W^2}C_0(x_i)\left[(1+\delta_2) (\sbar b)_{V-A}(\bar{\nu}\nu)_V
-(1+\delta_2)(\sbar b)_{V-A}(\bar{\nu} \nu)_A\right]
\end{align}
in which we have introduced two parameters  
\begin{equation}
\delta_1=\frac{s_W H_0(x_i)}{4c_W C_0(x_i)}
\left[\epsilon \frac{4c_W s_W}{1-4s_W^2}+\epsilon_Z
\right],\qquad
\delta_2=\epsilon_Z \frac{s_WH_0(x_i)}{4c_WC_0(x_i)}.
\end{equation}
Incoperating box diagram contribution, the
Wilson coefficients of operators
$Q_9^{\bar{\ell} \ell}=(\sbar b)_{V-A} (\bar{\ell}\ell)_V$, 
$Q_{10}^{\bar{\ell} \ell}=(\sbar b)_{V-A} (\bar{\ell}\ell)_A$
as well as quark-neutrino operator 
 $Q_{V-A}^{\bar{\nu}\nu}=(\sbar b)_{V-A} (\bar{\nu}\nu)_{V-A}$ 
can be explicitly extracted as 
\begin{align}
& C_9^{\bar{\ell}\ell}=\lambda_i \frac{G_F}{\sqrt{2}}\frac{\alpha}{2\pi s_W^2}
\Big[s_W^2\cdot 4\left(Z_0(x_i)+\Delta Z_0(x_i)\right)-\left( Y_0(x_i)+\Delta Y_V(x_i)\right)\Big]\nonumber\\
&C_{10}^{\bar{\ell}\ell}=\lambda_i\frac{G_F}{\sqrt{2}}\frac{\alpha}{2\pi s_W^2}
\Big[ Y_0(x_i)+\Delta Y_A(x_i)\Big]\\
&C^{\bar{\nu}\nu}= \lambda_i\frac{G_F}{\sqrt{2}}\frac{\alpha}{2\pi s_W^2}
\Big[ X_0(x_i)+\Delta X_0(x_i)\Big]\nonumber
\end{align}
in which the $X, Y, Z$ functions are given in the combination of
penguin and box diagrams
$X_0(x)=C_0(x)-4B_0(x),
Y_0(x)=C_0(x)-B_0(x),
Z_0(x)=C_0(x)+\frac14 D_0(x).
$
The corresponding corrections to $X, Y, Z$ are then given as
\begin{subequations}
\begin{align}
&\Delta Y_V(x)=\delta_1 C_0(x),\qquad
\Delta Z_0(x)=\delta_1 C_0(x),\\
&\Delta X_0(x)=\delta_2 C_0(x),\qquad
\Delta Y_A(x)=\delta_2 C_0(x).
\end{align}
\end{subequations}
Especially, we find that the modification to $Y$ has two types due to
different Dirac structure of lepton pair in the operators, while
modification to $X$ and $Z$ function are in a fixed way.
The modifications can be written in a more explicit form as
\begin{subequations}
\begin{align}
&\Delta X_0(x)=\frac14 t_W \epsilon_Z H_0(x)\\
&\Delta Y_A(x)=\frac14 t_W \epsilon_Z H_0(x)\\
&\Delta Y_V(x)=\left[\epsilon\frac{s_W^2}{1-4 s_W^2}
+\epsilon_Z \frac{t_W}{4}\right]{H_0(x)}\\
&\Delta Z_0(x)=\left[\epsilon\frac{s_W^2}{1-4 s_W^2}
+\epsilon_Z \frac{t_W}{4}\right]{H_0(x)}
\end{align}
\end{subequations}
In the limit of $\epsilon\to 0$, $\delta_1=\delta_2$, 
the modifications are identical. 
However, if $\epsilon_Z\to 0$ (exactly dark photon model case),
leading to  $\Delta X_0=\Delta Y_A=0$, then
the phenomenology is much more tedious.

\section{Phenomenology}

The physical observables can be classified into two types in dark $Z$ model.
One type relating to box diagram, like the mass difference of neutral meson, is not 
modified by $Z'$. 
The other one involving photon and $Z$ penguins,
such as the direct CP violation in $K\to\pi\pi$, does change.
In this section, we will choose several typical
processes to see a generic effect of $Z'$ on flavour physics.

\subsection{$B_q \to \mu^+\mu^-$}
It has been hoped for decades that NP might be unfolded  in rare decay $B_s\to \mu^+\mu^-$.
However, no hint of NP appeared in $B_s\to \mu^+\mu^-$ mode from LHC Run I data, given by
the full combination results of CMS and LHCb 
\cite{Bsmumu:CMS-LHCb} 
\begin{subequations}
\begin{align}
&\mathcal{B}(B_s\to \mu^+\mu^-)=(2.8^{+0.7}_{-0.6})\times 10^{-9}\\
&\mathcal{B}(B_d\to \mu^+\mu^-)=(3.9^{+1.6}_{-1.4})\times 10^{-10}.
\end{align}
\end{subequations}
Though $B_s\to \mu^+\mu^-$ turns out to be SM-like,
there remains a hope for NP in the much rarer mode $B_d\to \mu^+\mu^-$ (for example, see \cite{Hou:2013btm}).

Due to the precise measurement of  decay $B_s\to\ell^+\ell^-$ is now realistic,
one should consider the effect of sizeable
 width difference $\Delta \Gamma_s$ 
in $B_s^0$-$\overline{B}_s^0$ oscillation. The theoretical
formula 
 has to be corrected to compare with  measured branching ratio \cite{Bsmumu-NF1, Bsmumu-NF2}
 which is denoted with a bar,
\begin{equation}
{\mathcal{B}}(B_s^0\to \ell^+\ell^-)=\left[\frac{1-y_s^2}{1+\mathcal{A}_{\Delta \Gamma}^{\ell^+\ell^-} y_s}
\right]\overline{\mathcal{B}}(B_s^0\to \ell^+\ell^-)
\end{equation}
where $y_s\equiv \frac{\Delta \Gamma_s}{2\Gamma_s}\equiv \frac{\Gamma_L^{(L)}-\Gamma_L^{(s)}}{2\Gamma_s}$,
$\mathcal{A}_{\Delta \Gamma}^{\ell^+\ell^-}= \frac{R_H^{\ell^+\ell^-}-R_L^{\ell^+\ell^-}}{R_H^{\ell^+\ell^-}+R_L^{\ell^+\ell^-}}$.
It is known $\mathcal{A}_{\Delta \Gamma}^{\ell^+\ell^-}=1$ in SM  \cite{Bsmumu-NF2}, thus
\begin{equation}
{\mathcal{B}}(B_s^0\to \ell^+\ell^-)=\left(1-y_s^2
\right)\overline{\mathcal{B}}(B_s^0\to \ell^+\ell^-).\label{eq:BsmumuThEx}
\end{equation}
the latest estimation of parameter $y_s$ is $y_s=0.069\pm 0.006$ given in \cite{Bsmumu-ys}.
Note in the dark photon model, the relation of eq. (\ref{eq:BsmumuThEx}) does not change. While for 
the rarer $B_d\to \ell^+\ell^-$ decay, the effect from  oscillation in $B^0$-$\overline{B}^0$ can be neglected
thus we do not take this correction.

The (uncorrected) SM branching ratio of $B_q\to \ell^+\ell^-$ is induced by $Z$ penguin and hence
 depends on $Q_{10}$, (see ref. \cite{Bsmumu:Buras}). Now incorporating $Z'$-penguin, which
 gives a similar component as $Z$, leads to
 \begin{equation}
\mathcal{B}(B_q\to \ell^+\ell^-)=\tau(B_q)\frac{G_F^2}{\pi}\left(\frac{\alpha}{4\pi s_W^2}\right)^2
f_B^2m_{\ell}^2 m_B\sqrt{1-\frac{4 m_{\ell}^2}{m_B^2}} \eta_{\mathrm{eff}}^2\left|\lambda_t\Big(Y_0(x_t)+
\Delta Y_A(x_t)\Big)\right|^2
\end{equation}
with $\eta_{\mathrm{eff}}=0.9882\pm 0.0024$  which takes into account NNLO QCD correction
and NLO electroweak correction \cite{Bsmumu:Buras}. 
Apparently the reason why only $\Delta Y_A$ contributes,
is exact with the same reason why photon penguin contribution vanishes.

\subsection{$B\to K^{(*)}\mu^+ \mu^-$}
The quest for NP in $B\to K^{(*)}\mu^+ \mu^-$ has been performed for a long time.
In the beginning the zero crossing-point $q_0^2$  is of  the first priority, however, $q_0^2$ 
turns out to be compatible with SM prediction finally from the released LHC data.
The remaining possibility for NP in this mode, the $P_5'$ problem,  requires more data to confirm.
Meanwhile for the $B\to K\ell^+\ell^-$ channel, there is a NP hint, so called $R_K$ problem, 
 which violates lepton universality.

The theoretical study for this channel 
has been developed for around 30 years, including multi-loop calculation of 
Wilson coefficients at high energy.
The most matured theoretical treatment in low energy to the semileptonic decays are based on QCDF.
It is not necessary to repeat the whole long story in this work.  Instead, we would like to simply
focus on $C_9$ and $C_{10}$ to see how the data constrain the NP parameter space.
In dark $Z$ model, the modification of $C_9$ and $C_{10}$ due to the dark $Z$ contribution is
\begin{subequations}
\begin{align}
&\Delta C_9(x_i)= 4  \Delta Z_0(x_i)-\frac{\Delta Y_V(x_i)}{s_W^2}\\
&\Delta C_{10}(x_i)=\frac{\Delta Y_A(x_i)}{s_W^2}
\end{align}
\end{subequations}

Driven by new data many efforts have been made to extract the information hidden inside the two coefficients.
In \cite{C9-10-1} a model-independent fit was taken based on $B\to X_s\ell^+\ell^-, B \to X_s\gamma,
B\to K^*\gamma$ and 
 $B\to K^*\mu^+\mu^-$, we will adopt their constraints on $\Delta C_9$ and $\Delta C_{10}$  at $2 \sigma$
 \footnote{
Later in another independent analysis \cite{C9-10-2}, the  global fit combining 
$B\to X_s\ell^+\ell^-, B\to K^*\mu^+\mu^-$,
$B\to K\mu^+\mu^-$ and $B_s\to \mu^+\mu^-$ together obtained a similar constraint on
$C_{10}$.}
\begin{subequations}
\begin{align}
& -1.5<\mathrm{Re}\left(\Delta C_{9}\right)<1.2,\qquad -2.8<\mathrm{Im}\left(\Delta C_{9}\right)<2.8\\
& -1<\mathrm{Re}\left(\Delta C_{10}\right)<1.5,\qquad  -3<\mathrm{Im}\left(\Delta C_{10}\right)<3.
\end{align}
\end{subequations}
In our scenario, the NP only exists in the change of real part of $C_{9/10}$.

\subsection{$K_L\to \mu\bar{\mu}$}

The branching ratio of $K_L\to \mu\bar{\mu}$ contains LD  and SD contribution.
The calculation of LD contribution remains a challenge in theory. 
Combining a latest theoretical LD estimation and experimental bound \cite{KLmumu-bd2}
\footnote{The SM prediction is then
$\mathcal{B}(K_L\to \mu^+ \mu^-)_{\mathrm{SD}}=(0.79\pm0.12)\times10^{-9}$, the experimental
value in PDG is \\$\mathcal{B}(K_L\to \mu^+ \mu^-)_{\mathrm{exp.}}=(6.84\pm0.11)\times10^{-9}$.}, 
the constraint to SD is
\begin{equation}
\mathcal{B}(K_L\to \mu^+ \mu^-)_{\mathrm{SD}}\leq 2.5\times 10^{-9}
\end{equation}
The branching ratio from SD 
(see ref. \cite{Buras-KLmumu-new}) 
\begin{equation*}
\mathcal{B}(K_L\to \mu^+ \mu^-)_{\mathrm{SD}}=\kappa_\mu
\left[\frac{\mathrm{Re}(\lambda_c)}{|V_{us}|}P_c(Y_K)
+
\frac{\mathrm{Re}(\lambda_t)}{|V_{us}|^5}\eta_YY_0(x_t)\right]^2
\end{equation*}
with  $\lambda_q=V_{qs}^*V_{qd}  (q=c, t)$ and  
$\kappa_\mu=(2.009\pm 0.017)\times 10^{-9}\left(\frac{|V_{us}|}{0.225}\right)^8,
P_c(Y_K)=(0.115\pm 0.018)\left(\frac{0.225}{|V_{us}|}\right)^8$ \cite{KLmumu-SM}, 
 QCD correction factor  $\eta_Y=1.012$ \cite{epsilon:Buras13}, now
 is modified as
\begin{equation}
\mathcal{B}(K_L\to \mu^+ \mu^-)_{\mathrm{SD}}=\kappa_\mu
\left[\frac{\mathrm{Re}(\lambda_c)}{|V_{us}|}P_c(Y_K)
+
\frac{\mathrm{Re}(\lambda_t)}{|V_{us}|^5}\eta_Y\Big(Y_0(x_t)
+\Delta Y_A(x_t)\Big) \right]^2,
\end{equation}
by
including the dark $Z$ contribution.

\subsection{$\frac{\varepsilon'}{\varepsilon}$}

Historically two approaches, operator production expansion (OPE) method
and penguin-box expansion (PBE) method, are adopted for the study of
direct CP violation in $K\to\pi\pi$, which  involves all the QCD penguin and 
electroweak penguin in SM.
For the phenomenology study here,
we make use of the simple analytical formula
based on the PBE method\cite{Buchalla:1990qz}.
 By modifying
 relevant parts due to the dark $Z$ effect, 
 an updated formula\footnote{We should keep in mind that
 the original formula was obtained by comparing with two 
 methods in SM\cite{Buchalla:1990qz}. 
 A more serious formula in dark $Z$ model should  be given by repeating this work similarly 
 due to different types of $\Delta Y$, which is beyond the scope of current work. In the numerically
 study below, we will take $\Delta Y_V$ as an example. 
However, we will understand the exact form of $\Delta Y$
should keep the paper's conclusion.}
  to depicted $\varepsilon'/\varepsilon$ is 
\begin{equation}
\mathrm{Re}\frac{\varepsilon'}{\varepsilon}=a
\mathrm{Im}(\lambda_t)\cdot F(x_t)
\end{equation}
where $F(x)$ is given by
\begin{equation}
F(x)=P_0+P_X \left[X_0(x)+\Delta X(x)\right]+P_Y\left[ Y_0(x) 
+\Delta Y(x)\right]+ P_Z \left[Z_0(x)+\Delta Z\right] +P_E E_0(x)
\end{equation}
and the factor $a=0.92\pm 0.03$~\cite{Buras:2014sba}, which takes into account
the correction due to $\Delta I=5/2$ transitions~\cite{Cirigliano:2003nn}.
Note the dark $Z$ modifies most parts of SM $F$ function but keep
the gluon penguin vertex $E(x)$ unchanged.
The coefficients $P_i$ ($i=0,X,Y,Z,E$) are given in terms of $R_6, R_8$
\begin{equation}
P_i=r_0^{(0)}+r_i^{(6)} R_6+r_i^{(8)} R_8.
\end{equation}
We adopt their numerical values for $\alpha_s(M_Z)=0.1185$~\cite{PDG14} given in
Table 1 of Ref.~\cite{Buras:2014sba}.
For the nonperturbative parameters,
we adopt the value
%
$R_8 = 0.6, R_6=1.1$.
%
The former one is obtained from lattice~\cite{Blum:2012uk},
with the translation by Ref.~\cite{Buras:2014sba}.
But a  reliable lattice result for $R_6$ is still lack, here we choose 
$10\%$ deviation from large N result.
The experimental value with $1 \sigma$ error for $\varepsilon^\prime/\varepsilon$ is
\begin{align}
&\frac{\varepsilon^\prime}{\varepsilon}
 \simeq {\rm Re}\left( \frac{\varepsilon^\prime}{\varepsilon}\right)
= (1.66\pm 0.23)\times 10^{-3},
\end{align}
taken form PDG~\cite{PDG14}.

\section{Results and Discussion}

\begin{figure}[t!]
\begin{center}
{
 \includegraphics[width=100mm]{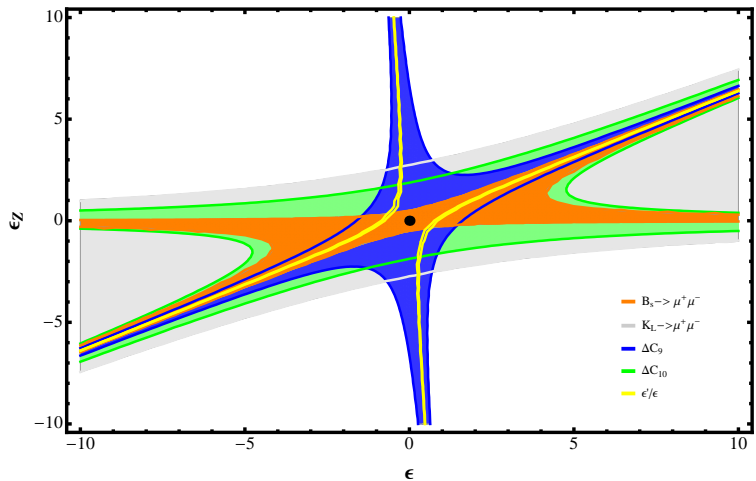}
}
\end{center}
\caption{
The allowed $\epsilon-\epsilon_Z$ parameter space by various experiments
if $m_{Z'}$ is less than $2m_{\mu}$ and hence ignored.
The meaning of colours are given as: grey stands for $K_L\to \mu^+\mu^-$, orange represents $B_s\to \mu^+\mu^-$,
green denotes $\Delta C_{10}$, blue stands for $\Delta C_{9}$ while
the yellow is one example of $\varepsilon'/\varepsilon$ with $(R_6, R_8)=(0.7, 1.1)$ and 
the black dot remarkes SM case.
  }\label{constraint}
\end{figure}

There are limited three free parameters in dark $Z$ model, $m_{Z'}, \epsilon$ and $\epsilon_Z$.
In our working scenario, the light $Z'$ mass is ignored thus it might be promising to
determine the remaining two by above observables.
We take the global fit of Wolfenstein parametrization of CKM matrix as input 
since $Z'$ does not change SM flavour structure.  
The other related input parameters have been given herebefore.
Combining $B_s\to \mu^+\mu^-$, $K_L\to \mu^+\mu^-$ and $\Delta C_9$, $\Delta C_{10}$
from a global fit of $B\to K^{(*)}\mu^+\mu^-$ and so on,
we plot allowed parameter space
in $(\epsilon,\epsilon_Z)$ plane shown in Fig.\ref{constraint}.

The ranges of $\epsilon$ and $\epsilon_Z$ are both shown in the section $(-10,10)$. 
Due to its large uncertainty SD of $K_L\to\mu^+\mu^-$, 
as presented in grey region, 
gives a pretty wide band with $\frac35$ slope.  
The green band with the same slope and less width of grey one, together 
with the  gap out of this band surrounded by a set of hyperbolic curve, is resulted from
$\Delta C_{10}$. Another parameter obtained from $B\to K^{(*)}\mu^+\mu^-$ samely, $\Delta C_9$,
constrains parameters in  blue colour, which
is restricted in two sets of hyperbolic curves and extended to slope
$\frac35$ direction as well as near vertical direction. Apparently
the latter part is excluded by $K_L\to\mu^+\mu^-$ (or $C_{10}$) while
the former one is embedding in green area.
Remarking in orange, the important $B_s \to \mu^+\mu^-$ also embeds in
$\Delta C_{10}$ but somehow has an overlap part with $\Delta C_9$.
In principle, $B_d\to\mu^+\mu^-$ can also be included. Considering its uncertainty is
larger than $B_s\to\mu^+\mu^-$, the allowed region is then also wider than  current
orange area thus we do not show it.
Here we also add in a typical constraint from  $\varepsilon'/\varepsilon$ by fixing 
its non-perturbative parameter $R_6$ and $R_8$. Showing as two narrow hyperbolic curves
in yellow, it looks $\varepsilon'/\varepsilon$ could give a very strict constraint.
However,  due to the large uncertainty of the two non-perturbative parameters,
especially $R_6$, one cannot take the direct CP violation of kaon too seriously. Nevertheless, once
$R_6$ is fixed within certain precisement in the future,  $\varepsilon'/\varepsilon$ 
can be taken as an important 
discrimination to further constrain the parameter. 

Aiming at the determination of parameter space of dark $Z$ model,  
we reach the
 allowed $(\epsilon, \epsilon_Z)$  in the narrow linear region
 \begin{equation}
 \epsilon_Z=\frac35 \epsilon
 \end{equation}
 and especially  $-2<\epsilon<2$, $-1<\epsilon_Z<1$ are favored. In other words,
 the order of mixing parameters is constrained to be $\mathcal{O}(1)$.
 Generally speaking, more FCNC processes can be considered to determine the bound,
 but we can believe $\mathcal{O}(1)$ should be the typical value from flavour physics.

 Many other works have already put  constraints on dark photon model.
 Though it is not exact same model as our working scenario, 
the obtained mixing angle in dark photon still enlightens parameters in 
dark $Z$ model.
 For example, based on supernova 1987A the limit on
 mixing angle could be $\mathcal{O}(10^{-12.5})$ for $m_{Z'}>2m_e$ \cite{Kazanas:2014mca}.
 A recent work from direct detection experiments such as XENON10 and XENON100, 
 for the absence of an ionization signal, puts a even more stringent limit on $\epsilon$  down 
 to $\mathcal{O}(10^{-15})$. One may expect if $\epsilon_Z$ is added, 
 the bound of $\epsilon$ will also be changed  from astrophysical observables and
 direct detection experiements.
 Meanwhile, the working scenario in flavour physics can also be modified to allow  $Z'$  resonance 
 production \cite{Davoudiasl:2012ag}, leading to a possible different constraint on mixing 
 parameters.
 In any case, given the bound from flavour physics in current scenario  $\mathcal{O}(1)$, 
 one may not expect a dramatical change (a more than $10$ orders of magnitude) happens
 unnaturally.

In the existence of dark $Z$, we have  investigated its effect in flavour physics especially by connecting
the effect to the newly measured processes at LHC. 
However, the obtained bound, $\mathcal{O}(1)$, may not compete 
with the corresponding one in traditional dark matter study.

\acknowledgments

FX would like to thank Dr. Xin-Qiang Li and Dr. Xiang-Dong Gao for the helpful discussion. This work is supported in part by the Ministry of Science and Technology of R.O.C. under Grant No. 104-2811-M-001-018 and National Natural Science Foundation of P.R.C. under Grant No. 11405074.


\appendix
\section{The calculation of $Z'$ penguin}
\label{sec:appendix}

The  effective $bsZ'$ vertex is in the form of
\begin{equation}
\Gamma_\mu(k_1,k_2)
=e\lambda_i \frac{g_2^2}{2(4\pi)^2 m_W^2}
\left[F_1(q_\mu\Slash q -q^2\gamma_\mu)\PL
+F_2 \,i\sigma_{\mu\nu}q^\nu m_b \PR\right],
\label{eq:FF}
\end{equation}
which satisfies Ward Identity approximately in the limit of $m_{Z'}\to 0$.
In the phenomenology study of this work, we only focus on $F_1$ contribution, which
has the relation with $H_0$ in section \ref{sec:Zprime}
\begin{equation}
H_0(x)=-2F_1(x).
\end{equation}
For convenience, in following calculation the effective vertex (\ref{eq:FF}) can be rewritten as
\begin{equation}
\Gamma^\mu(k_1,k_2)
=e\lambda_i \frac{g_2^2}{2(4\pi)^2 m_W^2}
\left[c_L T_L^\mu +c_b T_b^\mu+c_sT_s^\mu\right]
\end{equation}
by introducing $c_b=F_1+F_2,
c_s=F_2-F_1,
c_L=-F_2$ and $T_L^\mu=m_b^2\gamma^\mu\PL$,
$T_b^\mu=\Slash k_1 k_1^\mu\PL,
T_s^\mu=\Slash k_1 k_2^\mu\PL$.
Particularly
we have
\begin{equation}
F_1=\frac12 (c_b-c_s).
\end{equation}
In the calculation, the $F_1$ function may be projected to
\begin{equation}
F_1=\epsilon F_1|_\gamma + \epsilon_Z F_1|_Z
\end{equation}
Following we will evaluate $F_1$  component by component.

\subsection{The $\epsilon$ component}
The contribution of $\epsilon$ component is exactly same as $b\to s\gamma$ case.
Here we recalculate the photon penguin contribution. 
In Feynman-t' Hooft gauge, with the approximation $m_s=0, q^2=0$ and making use of quark on-shell condition
as well as unitary CKM relation, we have the total contribution as
\begin{equation}
i\Gamma^\mu\big|_\gamma=ie\lambda_i \frac{g_2^2}{2(4\pi)^2 }\sum_{j}\mathcal{M}_j^\mu\big|_\gamma
\end{equation}
with the individual amplitude from each diagram
\begin{align}
& \mathcal{M}_a\big|_\gamma= -\frac43 \Big[\left( -2C_{00}+m_t^2 C_0-m_b^2(C_0+2C_1+C_2+C_{11}+C_{12})\right)T_L
\nonumber\\
&\hspace{2cm} 
+2(C_{11}+C_1)T_b + 2(C_0+C_1+C_2+C_{12}) T_s\Big]\\
& \mathcal{M}_b\big|_\gamma=-\frac23 \frac{m_t^2}{m_W^2} \Big[ \left(-2C_{00}+\frac12+m_t^2 C_0
-m_b^2(C_{11}+C_1+C_{12}+{C_0})\right)T_L\nonumber\nonumber\\
&\hspace{2cm}
+2C_{11}T_b +2(C_{12}{-C_2})T_s\Big]\nonumber\\
&\mathcal{M}_c\big|_\gamma=-\Big[\left(12 C_{00}+m_b^2(3C_1+2C_2+2C_{11}+2C_{12})\right)T_L\nonumber\\
&\hspace{2cm}
+ (4C_{11}+2C_1) T_b +(4C_{21}-2C_1-2C_2)T_s\Big]\nonumber\\
&\mathcal{M}_d\big|_\gamma=-\frac{m_t^2}{m_W^2}\Big[ 2C_{00} T_L + (2C_{11}+3C_1{+C_0})T_b + (2C_{21}+C_1
{+2C_2+C_0})T_s\Big]\nonumber\\
& \mathcal{M}_e\big|_\gamma=m_t^2 C_0 T_L\nonumber\\
& \mathcal{M}_f\big|_\gamma=(m_t^2 C_0{+m_b^2 C_1}) T_L{+2C_2 T_s}\nonumber\\
& \mathcal{M}_{g}\big|_\gamma=\frac13\left[(B_0+B_1)\left(2+\frac{m_t^2}{m_W^2}\right) {-\frac{m_t^2}{m_W^2}\left(
B_0(m_b^2,m_W^2,m_t^2)-B_0(0,m_W^2,m_t^2)
\right)}
\right]T_L\nonumber
\end{align}
in which $B_i, C_j$  ($i=0,1; j=0,1,2,11,12,21,22$) are Pasarrino-Veltman integrals\cite{'tHooft:1978xw}, 
and the positions for variables 
 are assigned as
 $B(m_b^2, m_W^2, m_t^2)$,  $C_{a,b}(m_b^2,0,0,m_W^2,m_t^2,m_t^2)$ for $a,b$ case, and
$C_{c,d,e,f}(m_b^2,0,0,m_t^2,m_W^2,m_W^2)$ for $c,d,e,f$ case. Note in the calculation, 
the light down type quark contribution in 
Goldstone-quark-quark vertex cannot be neglected.

The PV functions can be reduced into basic scalar function $B_0, C_0$, 
and we further perform Taylar expansion up to $m_b^4$.
Sum up all the contribution together, we have
\begin{subequations}
\begin{align}
&i\Gamma^\mu\big|_\gamma=ie\lambda_i \frac{g_2^2}{2(4\pi)^2 m_W^2}\left[
\left(-\frac13-2\Delta_\epsilon\right)\frac{T_L}{m_b^2}+c_L  T_L
+c_b T_b + c_s T_s
\right]\label{vertex:Gamma}\\
& c_L\big|_\gamma=\frac{x_t^2(2-3x_t)\ln x_t}{2(x_t-1)^4}-\frac{22 x_t^3-153 x_t^2+159x_t -46}{36(x_t-1)^3}\\
& c_b\big|_\gamma=\frac{(3x_t^4-3x_t^3+36x_t^2-32x_t+8)\ln x_t}{18(x_t-1)^4}+
\frac{19 x_t^3-222 x_t^2+165x_t-34}{108(x_t-1)^3}\\
&c_s\big|_\gamma=\frac{(-3x_t^4+57x_t^3-72x_t^2+32x_t-8)\ln x_t}{18(x_t-1)^4}+
\frac{113x_t^3-696x_t^2+789x_t-242}{108(x_t-1)^3}
\end{align}
\end{subequations}
The first term in eq. (\ref{vertex:Gamma}) gives zero contribution
after applying  the unitary triangle relation $\sum_i \lambda_i \cdot \mathrm{constant}=0$.
Also we have
\begin{eqnarray}
c_b\big|_\gamma-c_s\big|_\gamma
&=&\frac{(3x_t^4-30x_t^3+54x_t^2-32x_t+8)\ln x_t}{9(x_t-1)^4}+\frac{-47x_t^3+237x_t^2-312x_t+104}{54(x_t-1)^3}
\nonumber\\
&=&-2\left(D_0(x_t)+\frac{26}{27}\right)
\end{eqnarray}
\begin{equation}
F_1|_\gamma = -D_0(x_i)
\end{equation}
in which $D_0(x_i)$ is vertex function of virtual photon in $b\to s \gamma$.
Note in my calculation, the $D_0, D'_0$ differs from the one in \cite{Buras:1998raa}, up to a minus sign, for
the different convention in QED vertex.

\subsection{The $\epsilon_Z$ component}

The total contribution for $b \to s Z$ is 
\begin{equation}
i\Gamma^\mu\big|_Z =i \frac{g_2^2}{2(4\pi)^2}\frac{g_2}{c_W}\lambda_i \sum_j \Gamma^\mu_{j}\big|_Z
\end{equation}
Compared with $b\to s \gamma$ case,
we only need to recalculate Fig.(a) and Fig.(b), 
\begin{subequations}
\begin{align}
\Gamma^\mu_{a}\big|_Z=& \frac23 s_W^2\cdot 2\Big\{ 
[-2C_{00}+m_t^2 C_0 -m_b^2(C_{11}+C_{12}+2C_1+C_2+C_0)]T_L^\mu\nonumber\\
&\hspace{2cm} +2(C_{11}+C_1)T_b^\mu + 2(C_{12}+C_2+C_1+C_0)T_s^\mu\Big\}\nonumber\\
&-\frac12\cdot 2\Big\{ 
[-2C_{00} -m_b^2(C_{11}+C_{12}+2C_1+C_2+C_0)]T_L^\mu\nonumber\\
&\hspace{2cm} +2(C_{11}+C_1)T_b^\mu + 2(C_{12}+C_2+C_1+C_0)T_s^\mu\Big\}\\
\Gamma^\mu_{b}\big|_Z=& \frac23 s_W^2\cdot \frac{m_t^2}{m_W^2}\Big\{ 
[-2C_{00}+\frac12 +m_t^2 C_0 -m_b^2(C_{11}+C_{12}+C_1+C_0)]T_L^\mu\nonumber\\
&\hspace{2cm} +2C_{11} T_b^\mu + 2(C_{12}-C_2)T_s^\mu\Big\}\nonumber\\
&-\frac12\cdot \frac{m_t^2}{m_W^2}\Big\{ 
[m_t^2 C_0 -m_b^2(C_{0}+C_{1})]T_L^\mu- 2C_2 T_s^\mu\Big\}
\end{align}
\end{subequations}
while the other contributions are obtained by the following
replacement
\begin{align}
& \mathrm{Fig.} (c): \quad e\rightarrow g_2 c_W\nonumber\\
& \mathrm{Fig.} (d): \quad e\rightarrow g_2 \frac{1-2s_W^2}{2c_W}\nonumber\\
& \mathrm{Fig.} (e),(f) : \quad e\rightarrow -\frac{g_2s_W^2}{c_W}\nonumber\\
& \mathrm{Fig.} (g): -Q_b \rightarrow -\left(-\frac12+\frac13 s_W^2\right)
\end{align}
with the same  PV function convention  as  in $b\to s\gamma$ case.


In this work, since we only consider the $(q_\mu\Slash q-q^2\gamma_\mu)$ term, thus only
the coefficients of
$T_b^\mu$ and $T_s^\mu$ are of interest. 
Especially they are listed individually,
\begin{subequations}
\begin{align}
m_W^2\Gamma^\mu_{a}\big|_Z=
& \left[\frac{-5x_t^2+22x_t-5}{18(x_t-1)^3}+\frac{(1-3x_t)\ln x_t}{3(x_t-1)^4}
+s_W^2\left(\frac{2(5x_t^2-22x_t+5)}{27(x_t-1)^3}+\frac{4(3x_t-1)\ln x_t}{9(x_t-1)^4}\right)
\right] T_b^\mu\nonumber\\
&+ \left[ \frac{20x_t^2-7x_t-7}{18(x_t-1)^3}-\frac{(6x_t^2-6x_t+1)\ln x_t}{3(x_t-1)^4}
\right.\nonumber\\
&\hspace{1cm}\left.
+s_W^2
\left(-\frac{2(20x_t^2-7x_t-7)}{27(x_t-1)^3}+
\frac{4(6x_t^2-6x_t+1)\ln x_t}{9(x_t-1)^4}\right)
\right]T_s^\mu\nonumber\\
m_W^2\Gamma^\mu_b\big|_Z=
& s_W^2\left(\frac{2(2x_t^2-7x_t+11)}{27(1-x_t)^3}+\frac{4x_t\ln x_t}{9(x_t-1)^4}\right)
 T_b^\mu\nonumber\\
&+ \left[ \frac{x_t(x_t-3)}{4(x_t-1)^2}+\frac{x_t\ln x_t}{2(x_t-1)^3}
\right.\nonumber\\
&\hspace{1cm}\left.
+s_W^2
\left(\frac{x_t(11x_t^2-43x_t+38)}{27(x_t-1)^3}-
\frac{2(3x_t-4)\ln x_t}{9(x_t-1)^4}\right)
\right]T_s^\mu\nonumber\\
m_W^2 \Gamma^\mu_{c}\big|_Z=
& \left[ \frac{-17x_t^2-8x_t+1}{18(x_t-1)^3}+\frac{x_t^2(x_t+3)\ln x_t}{3(x_t-1)^4}
+s_W^2\left( \frac{17x_t^2 + 8x_t-1}{18(x_t-1)^3}-\frac{x_t^2(x_t+3)\ln x_t }{3(x_t-1)^4}\right)\right]T_b^\mu\nonumber\\
& + \left[ \frac{-38x_t^2+43x_t-11}{9(x_t-1)^3}+\frac{2x_t^2(4x_t-3)\ln x_t}{3(x_t-1)^4}\right.\nonumber\\
&\left.\hspace{0.5cm}
+s_W^2\left(-\frac{2x_t^2(4x_t-3)\ln x_t}{3(x_t-1)^4}
+\frac{38x_t^3-43x_t+11}{9(x_t-1)^3}\right)\right]T_s^\mu\nonumber
\end{align}
\end{subequations}
\begin{subequations}
\begin{align}
m_W^2 \Gamma^\mu_d\big|_Z=
&\left[ \frac{x_t(x_t^2-8x_t-17)}{72(x_t-1)^3}+\frac{x_t^2(x_t^2-3x_t+6)\ln x_t}{12(x_t-1)^4}\right.\nonumber\\
&\hspace{0.5cm}\left.
+s_W^2\left(\frac{x_t(x_t^2-8x_t-17)}{36(1-x_t)^3}
-\frac{x_t^2(x_t^2-3x_t+6)\ln x_t}{6(x_t-1)^4}\right)\right]T_b^\mu\nonumber\\
&+\left[ \frac{x_t(23x_t^2-22x_t-13)}{72(x_t-1)^3}-\frac{(x_t^2+3x_t-6)\ln x_t}{12(x_t-1)^4}\right.\nonumber\\
&+\hspace{0.5cm}
\left. s_W^2\left( \frac{x_t(-23x_t^2+22x_t+13)}{36(x_t-1)^3}+
\frac{x_t^2(x_t^2+3x_t-6)\ln x_t}{6(x_t-1)^4}\right)\right]T_s^\mu\nonumber\\
m_W^2 \Gamma^\mu_e\big|_Z=&0\nonumber\\
m_W^2 \Gamma^\mu_f\big|_Z=
&s_W^2 \left( -\frac{1-3x_t}{2(x_t-1)^2}-\frac{x_t^2\ln x_t}{(x_t-1)^3}\right)T_s^\mu\nonumber\\
m_W^2 \Gamma^\mu_g\big|_Z=&0\nonumber
\end{align}
\end{subequations}

Sum up together, we have
\begin{eqnarray}
c_b\big|_Z&=&\left[ \frac{-35x_t^3+156x_t^2-213x_t+20}{216(x_t-1)^3}+
\frac{(-3x_t^4+9x_t^3-18x_t^2+28x_t-4)\ln x_t}{36(x_t-1)^4}\right.\\
&&\left.\hspace{0.5cm}
+c_W^2\left(\frac{19x_t^3-222x_t^2+165x_t-34}{108(x_t-1)^3}
+\frac{(3x_t^4-3x_t^3+36x_t^2-32x_t+8)\ln x_t}{18(x_t-1)^4}\right)\right]\nonumber
\end{eqnarray}
\begin{eqnarray}
c_s\big|_Z&=&\left[ \frac{-103x_t^3+438x_t^2 -507x_t +136}{216(x_t-1)^3}
+\frac{(3x_t^4-27x_t^3+36x_t^2-10x_t+4)\ln x_t}{36(x_t-1)^4}\right.\\
&& \hspace{0.5cm}\left.
+c_W^2\left( \frac{113x_t^3-696x_t^2+789x_t-242}{108(x_t-1)^3}
+\frac{(-3x_t^4+57x_t^3-72x_t^2+32x_t-8)\ln x_t}{18(x_t-1)^4}\right)\right]\nonumber
\end{eqnarray}
Now we obtain the $\tilde{D}_0$ function as
\begin{eqnarray}
c_b\big|_Z-c_s\big|_Z
&=&2\left[ \frac{34x_t^3-141x_t^2+147x_t-58}{216(x_t-1)^3}
+\frac{(-3x_t^4+18x_t^3-27x_t^2+19x_t-4)\ln x_t}{36(x_t-1)^4}\right.\nonumber\\
&&\left. + c_W^2\left( \frac{-47x_t^3+237x_t^2-312x_t +104}{108(x_t-1)^3}
+\frac{(3x_t^4-30x_t^3+54 x_t^2-32 x_t+8)\ln x_t}{18(x_t-1)^4}\right)\right].\nonumber
\end{eqnarray}
\begin{equation}
\tilde{D}_0(x_i)\equiv -F_1|_Z=\frac{1}{2s_W c_W}\left(
c_b\big|_Z-c_s\big|_Z
\right)
\end{equation}
Take into account $D_0$ and $\tilde{D}_0$ together, the vertex function for $Z'$ penguin $H_0$ is produced. 



\end{document}